\documentclass[prd,twocolumn,superscriptaddress, nofootinbib, notitlepage, 10pt]{revtex4-1}  
\usepackage{graphicx}
\usepackage{amsmath}
\usepackage{amsfonts}
\usepackage{amssymb}
\usepackage{appendix}
\usepackage{mathtools}
\usepackage{comment}
\usepackage{color}
\usepackage{slashed}
\usepackage{hyperref}
\usepackage{subfigure}
\usepackage{setspace}
\usepackage{enumitem}
\usepackage{longtable}
\usepackage{wasysym}
\usepackage[usenames,dvipsnames]{xcolor}
\usepackage{bm}
\usepackage[letterpaper, margin=.8in, top=0.85in, bottom=0.85in]{geometry}

\newcommand{\Z}{\mathbb{Z}}
\newcommand{\Lagr}{\mathcal{L}}

\begin{document}

\singlespacing

\title{Baryon Number, Lepton Number, and Operator Dimension in the SMEFT
	with Flavor Symmetries}
	\author{Andreas Helset}
	\affiliation{Niels Bohr International Academy and Discovery Center, Niels Bohr Institute, University of Copenhagen, 
	Blegdamsvej 17, DK-2100 Copenhagen \O, Denmark}
	\author{Andrew Kobach}
	\affiliation{Department of Physics, University of California, San Diego, La Jolla, CA 92093, USA}

\begin{abstract}
For a large set of flavor symmetries, the lowest-dimensional baryon- or lepton-number violating 
operators in the Standard Model effective field theory (SMEFT) with flavor symmetry 
are of mass dimension 9.
As a consequence, baryon- and lepton-number violating processes are further suppressed with
the introduction of flavor symmetries, e.g.,
the allowed scale associated with proton decay is typically lowered to 
$10^5~$GeV, which is significantly lower than the GUT scale. 
To illustrate these features, we discuss Minimal Flavor Violation
for the Standard Model augmented by sterile neutrinos. 
\end{abstract}

\date{\today}

\maketitle

\section*{Introduction}

The Standard Model (SM) has been spectacularly successful at describing 
interactions among the known elementary particles. However, it suffers from some shortcomings.
An incomplete list of phenomena not fully explained by the SM could include
the experimental evidence for the existence of dark matter 
\cite{Zwicky:1933gu,Blumenthal:1984bp,Einasto:2009zd,Angle:2011th,Strigari:2013iaa}, 
the observational fact of 
a global baryon asymmetry in the universe \cite{Spergel:2003cb,Komatsu:2008hk,Hinshaw:2008kr,Komatsu:2010fb,Ade:2013sjv,Ade:2015xua}, and evidence of non-zero mass terms for 
(at least two generations of) the neutrinos from observed neutrino oscillations \cite{Fogli:2012ua,Schwetz:2011qt,Forero:2014bxa,Gonzalez-Garcia:2015qrr,Esteban:2018azc}.

Given that the SM is incomplete, insofar as it is unable to explain the 
above mentioned experimental facts, and with no conclusive hint from collider experiments
of new particles beyond those in the SM at the TeV scale,
the scale of new physics could be much higher.
If so, the SM can be extended in an agnostic, model-independent approach to an effective field theory, namely the 
Standard Model Effective Field Theory (SMEFT).
The SMEFT is constructed by adding a complete set of higher-dimensional operators, which give rise
to independent $S$-matrix elements, built out of the SM 
field content, respecting the underlying local $SU(3)_c \otimes SU(2)_L \otimes U(1)_Y$ gauge symmetry 
\cite{Buchmuller:1985jz,Grzadkowski:2010es}.

Baryon number and lepton number are
accidental symmetries of the SM. Thus, within the confines of the SM, 
the baryon asymmetry in the universe could
only come about through non-perturbative processes, e.g., high-temperature
sphaleron processes \cite{Kuzmin:1985mm}. Baryon-number violating processes are 
exponentially suppressed at low temperatures \cite{tHooft:1976rip}.
Whether lepton number is conserved or not is intimately related to neutrino masses, and, in particular, if neutrinos are Majorana fermions, this may point to the existence of an additional scale above the weak scale.

One new feature that occurs in the SMEFT is that baryon number and lepton number can be 
violated by higher-dimensional operators.
Lepton number and baryon number are always integers. For lepton number, this follows directly
from the definition, while it is a consequence of hypercharge invariance in the case 
of baryon number. 
There is a close connection between lepton number, baryon number, and the mass dimension of
operators \cite{Kobach:2016ami};
\begin{align}
	\label{eq:mod2}
	\frac{\Delta B - \Delta L}{2} \equiv d \mod 2,
\end{align}
where $\Delta B$ is the baryon number, $\Delta L$ is the lepton number and  $d$ is the mass dimension of the operator. Some consequences of Eq.~\eqref{eq:mod2},
among many others, are that $(\Delta B - \Delta L)/2$ must be an integer, and 
no operator with odd mass dimension can preserve both baryon number and lepton number.
See Ref.~\cite{Kobach:2016ami} for more details.

Consider the lowest-dimensional operators that violate baryon number and/or lepton number.
The only effective operator at dimension 5 is the famous Weinberg operator \cite{Weinberg:1979sa}.
This operator violates lepton number by two units, $|\Delta L|=2$,
and is associated with 
Majorana mass terms for the neutrinos below the electroweak scale,
generated by, for example, the seesaw mechanism \cite{Minkowski:1977sc}.
At dimension 6, all operators satisfy $\Delta B - \Delta L = 0$.
Many of the effective operators preserve both baryon number and lepton number,
$\Delta B = \Delta L = 0$. There are also some operators consisting of four fermion fields 
of the form $qqql$, where $q$ is a generic quark field and $l$ is a generic lepton field.
These operators violate baryon number and lepton number by $\Delta B = \Delta L = \pm 1$.
As a consequence, they can mediate proton decay, through the dominant two-body
decay $p\rightarrow M l$, where $p$ is the proton and $M$ is a meson.
The experimental null result for such decay processes has pushed the 
allowed scale for 
baryon-number violating
operators with mass dimension 6 to around $10^{15}$ GeV \cite{Nishino:2012bnw}.

Eq.~\eqref{eq:mod2} and the results in Ref.~\cite{Kobach:2016ami} apply to the SMEFT, with one generation of fermions. They 
remain true with the extension to multiple generations, and with the inclusion of 
flavor symmetries. 
A flavor symmetry is a global symmetry among the generations of fermions.
We will consider what happens to the general relation between baryon number, lepton number,
and operator dimension when flavor symmetries are present. 

The paper is organized as follows. 
We start by discussing the general relation between baryon number, lepton number and 
operator dimension, following from the local symmetries and field content of the 
SM.
Then, we discuss how this relation gets modified when certain flavor symmetries are present,
and discuss in detail an explicit example of Minimal Flavor Violation (MFV) for the 
SM augmented
by sterile neutrinos. 
We end by discussing implications for proton decay and Majorana neutrino masses.

\section*{$\Delta B$, $\Delta L$ and operator dimension}
The SM field content consists of the fermions $\{ L, e^c, Q, u^c, d^c\}$ and the 
Hermitian conjugate fields, the gauge
bosons for the gauge groups $SU(3)_c \otimes SU(2)_L \otimes U(1)_Y$, and the Higgs boson $H$.
The fermions are in the respective representations of the gauge groups as 
\begin{align}
	&Q \sim (\bm 3, \bm 2)_{1/6}, \quad u^c \sim (\bm{\overline{3}}, \bm 1)_{-2/3}, \quad
	d^c \sim (\bm{\overline{3}}, \bm 1)_{1/3}, \nonumber \\
	&L \sim (\bm 1, \bm 2)_{-1/2}, \quad  e^c \sim (\bm 1, \bm 1)_1, \quad 
	\nu^c \sim (\bm 1, \bm 1)_0,
\end{align}
where we have also included a sterile neutrino $\nu^c$ for generality.
All the fermions are the in the $(\bm 2,\bm 1)$ representation of the Lorentz group $SU(2)_L\otimes SU(2)_R$,
with the Hermitian conjugate fields being in the $(\bm 1, \bm 2)$ representation.
The Higgs field is in the representation 
\begin{align}
	H \sim (\bm 1, \bm 2)_{1/2}
\end{align}
of the gauge groups and is a Lorentz singlet.

\begin{figure}
	\includegraphics[width=\linewidth]{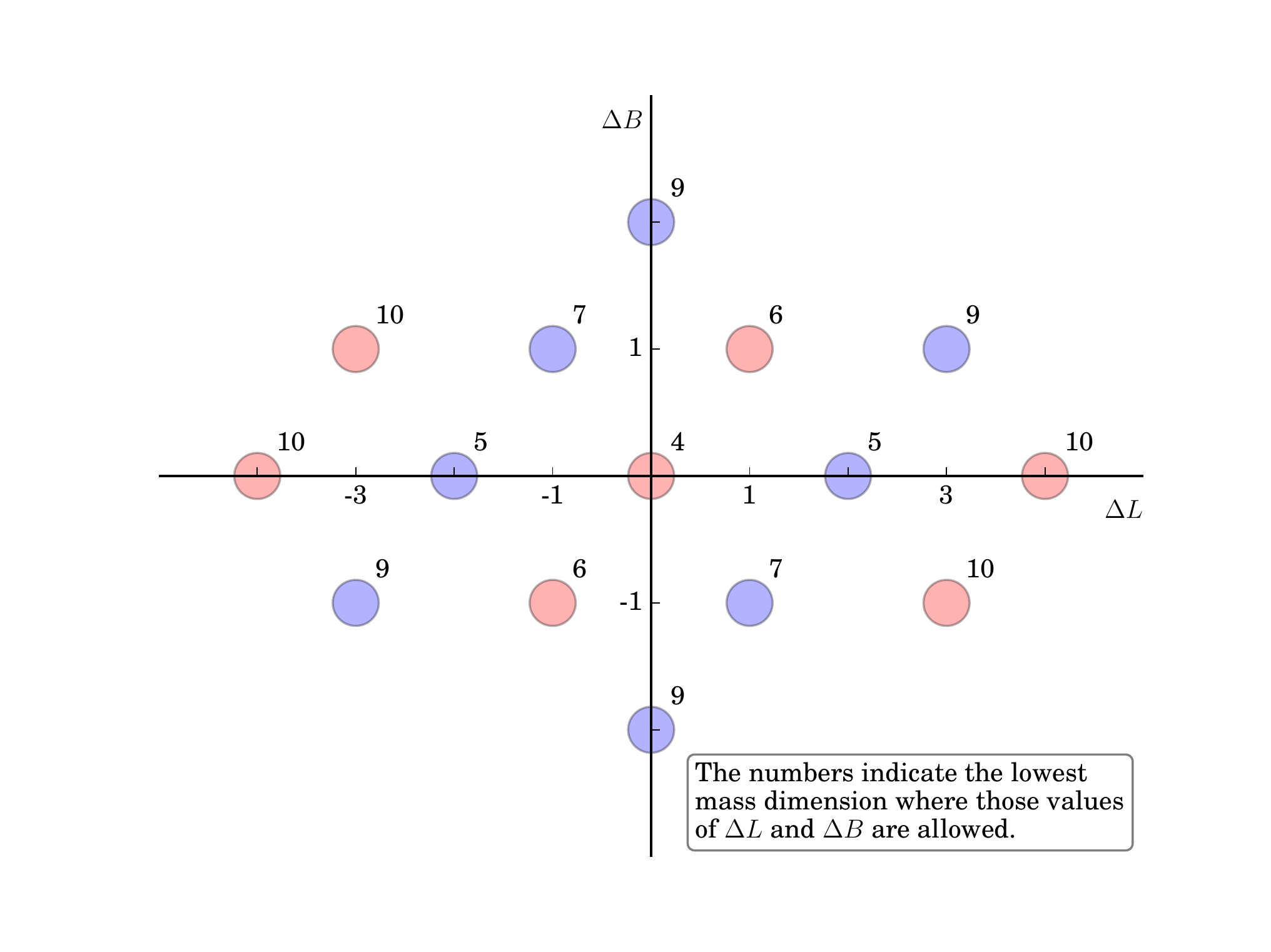}
	\caption{The $(\Delta L,\Delta B)$ values of operators with different 
		mass dimension $d$ without any flavor symmetry. The numbers indicate 
	the lowest mass dimension where the $(\Delta L,\Delta B)$ value is allowed.
Even (odd) dimensional operators are shown in red (blue).}
	\label{fig:noFlavor}
\end{figure}

We will denote the number of various fermion fields and their Hermitian conjugate fields 
in an operator 
by e.g., $N_e$ and $N_{e^\dagger}$ for the fermion fields $e^c$ and $e^{c\dagger}$ etc. 
Baryon number and lepton number are defined as
\begin{align}
	\label{eq:DeltaB}
	&\Delta B \equiv \frac{1}{3}\left( N_Q + N_{u^\dagger} + N_{d^\dagger} \right)
	- \frac{1}{3}\left( N_{Q^\dagger} + N_u + N_d \right), \\
	\label{eq:DeltaL}
	&\Delta L \equiv \left( N_L + N_{e^\dagger} + N_{\nu^\dagger} \right)
	- \left( N_{L^\dagger} + N_e + N_\nu \right).
\end{align}
Both baryon number and lepton number are integers, $\Delta B \in \Z$ and $\Delta L \in \Z$.
For lepton number, this can be seen directly from Eq.~\eqref{eq:DeltaL}. For baryon number,
this follows from hypercharge invariance \cite{Kobach:2016ami}.
As the fermion fields have mass dimension $3/2$, it follows directly that the mass dimension
of an operator which violates baryon number and/or lepton number is bounded by
\begin{align}
	\label{eq:dmin}
	d_{\textrm{min}} \geq \frac{9}{2}|\Delta B| + \frac{3}{2}|\Delta L|.
\end{align}
By combining Eq.~\eqref{eq:dmin} with Eq.~\eqref{eq:mod2}, we show the allowed $(\Delta L,\Delta B)$ 
values for various mass dimensions of operators in Fig.~\ref{fig:noFlavor}, 
where the sterile neutrinos are excluded.

\section*{Flavor Symmetry}
We consider the allowed baryon number and lepton number values of the higher-dimensional
operators when the fermions in the SM transform non-trivially under a 
continuous flavor group $G_F$.\footnote{This requirement excludes some prominent flavor models,
	see e.g. Refs.~\cite{Ma:2001dn,Altarelli:2010gt,Ishimori:2010au}.}
As the quarks and leptons are charged differently under the $SU(3)_c$ gauge group, we let the flavor group $G_F$ be factorized into a direct product of two distinct flavor groups,
one for the quarks, $G_q$, and one for the leptons, $G_l$,
\begin{align}
	G_F = G_q \otimes G_l.
\end{align}	
Also, we let all the generations of the fermions be charged democratically, i.e.,
they form irreducible representations of the flavor group.

The Yukawa terms will in general break the flavor symmetry \cite{Yukawa:1935xg}. 
The flavor symmetry can formally be restored by promoting the Yukawa couplings to spurion 
fields, transforming 
appropriately
to form invariants under the flavor group.
Spurion fields are auxiliary fields with non-trivial transformation properties, but are 
not part of the Fock space, i.e., they do not contribute to the $S$-matrix.

In order to form operators which are singlets under the flavor group, more than one 
flavor multiplet is required (or none). 
The constraint that a single quark field cannot appear in an operator is already encoded in the 
$SU(3)_c$ invariance. The leptons, however, have no such constraint. Thus, imposing a 
flavor symmetry restricts the leptons to
\begin{align}
	\label{eq:Nlnot1}
	N_L+N_{L\dagger}+N_e+N_{e^\dagger} \neq 1,
\end{align}
where we again have excluded the sterile neutrinos.
This basic fact severely restricts the allowed values of baryon and lepton number.
By combining Eqs.~\eqref{eq:dmin} and~\eqref{eq:Nlnot1}, the mass dimension of a baryon- or
lepton-number violating operator is bounded by
\begin{align}
	\label{eq:dminflavor}
	d_{\textrm{min}} \geq \frac{9}{2}|\Delta B|
	+ \frac{3}{2}|\Delta L| + 3 \delta_{|\Delta L|,1},
\end{align}
where $\delta_{|\Delta L|,1}$ is the Kronecker delta.
From Eqs.~\eqref{eq:mod2} and~\eqref{eq:dminflavor}, we find that no baryon- or lepton-number violating
operator is allowed below dimension 9, except for $|\Delta L| = 2$.

This constraint excludes the dimension-6 operators $qqql$. The quarks transform under
the quark flavor group and could form a singlet. However, the lepton field by itself,
being in a non-trivial representation of the lepton flavor group,
cannot form a singlet under the lepton flavor group by itself. Thus,
the operators break the flavor symmetry.
Implications for proton decay are discussed later.

The Weinberg operator also is forbidden, with one notable exception. 
The operator consists of two lepton fields in the same representation, 
and the requirement that the lepton fields transform under the lepton flavor group as a triplet excludes this operator.
The only case where two triplets could form a singlet is the case where the symmetry group is
$G_l=SU(2)$ (not to be confused with the electroweak gauge group $SU(2)_L$), and the lepton fields are 
in the adjoint representation \cite{Wilczek:1978xi}.

We now want to see which $(\Delta L,\Delta B)$ values are allowed with the inclusion of a flavor symmetry.
Let us start with the quarks. By letting them transform as triplets under the flavor group,
a necessary (but not sufficient) requirement of an operator being invariant under the quark flavor group and the $SU(3)_c$ gauge group is
	\begin{align}
		\frac{1}{3}\left( N_Q + N_{u^\dagger} + N_{d^\dagger} \right)
		- \frac{1}{3} \left( N_{Q^\dagger} + N_u + N_d \right) \in \Z.
	\end{align}
This is nothing but the result that baryon number takes integer values.

\begin{figure}
	\includegraphics[width=\linewidth]{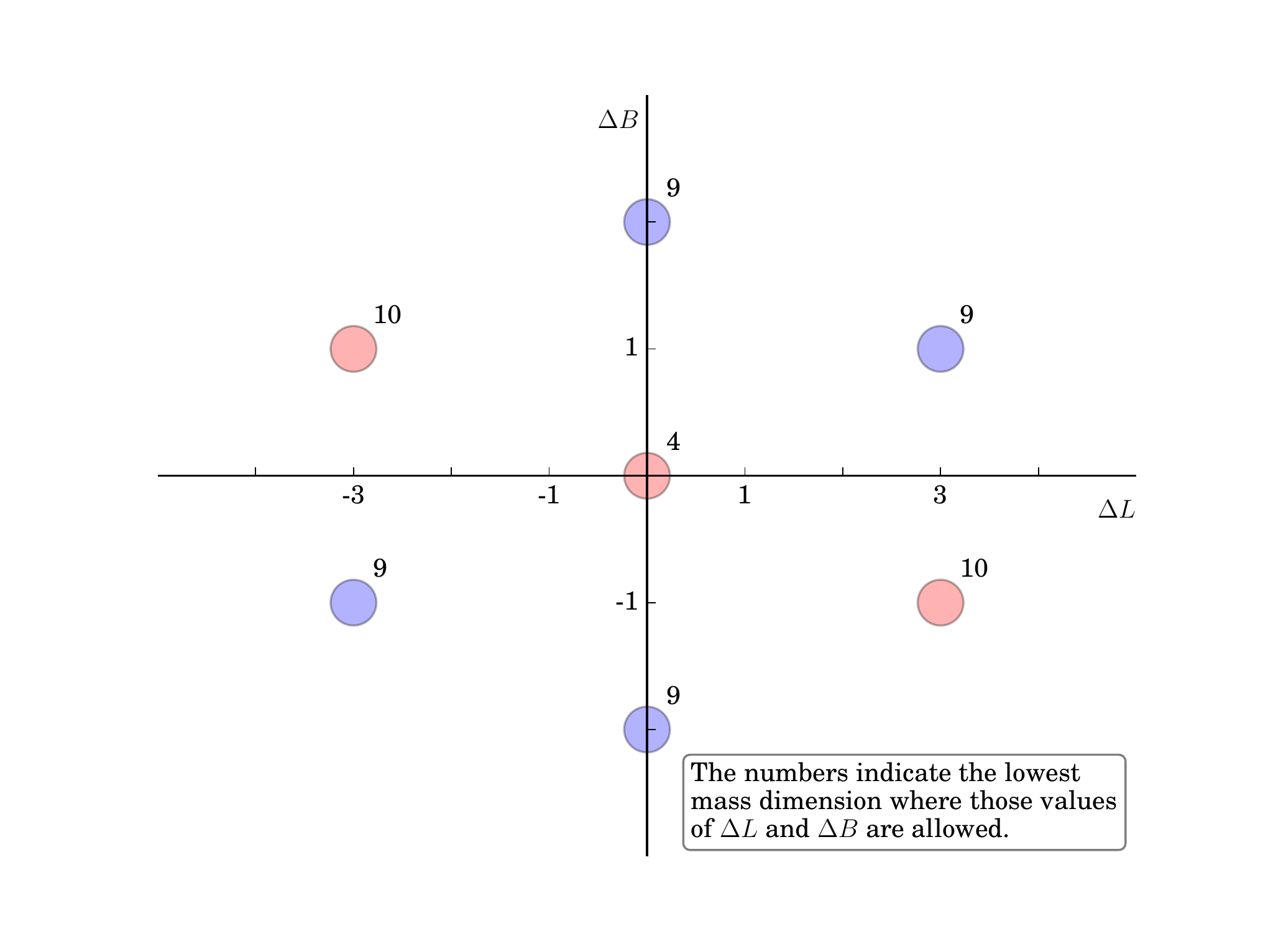}
	\caption{The $(\Delta L,\Delta B)$ values of operators with different 
		mass dimension $d$ with flavor symmetry, where the leptons are not 
		in the adjoint representation of an $SU(2)$ flavor group. The numbers indicate 
	the lowest mass dimension where the $(\Delta L,\Delta B)$ value is allowed.
Even (odd) dimensional operators are shown in red (blue).}
	\label{fig:Flavor}
\end{figure}

Consider the case where the leptons are in the fundamental representation of an $SU(3)$ flavor group, and not in the adjoint representation of an $SU(2)$ flavor group.
They must form invariants subject to the constraint 
\begin{align}
	\frac{1}{3}\left( N_L + N_{e^\dagger} \right)
	- \frac{1}{3}\left( N_{L^\dagger} + N_e \right) \in \Z.
\end{align}
From the definition of lepton number, Eq.~\eqref{eq:DeltaL}, and with no sterile neutrinos,
we have that
\begin{align}
	\label{eq:DeltaL3}
	\frac{1}{3}\Delta L \in \Z.
\end{align}
Lepton number can only be violated in multiples of 3. From this we can immediately see that 
no Majorana mass term is allowed.
We show the allowed baryon number and lepton number values of the operator basis with flavor symmetry
in Fig.~\ref{fig:Flavor}.
The baryon- or lepton-number violating operators with the lowest mass dimension have 
mass dimension 9.
By comparing Figs.~\ref{fig:noFlavor} and \ref{fig:Flavor}, we see that the set of allowed baryon number and lepton number 
values has been severely restricted.

\section*{Minimal Flavor Violation}
We now turn to an explicit example of a flavor symmetry, namely Minimal Flavor Violation (MFV) \cite{DAmbrosio:2002vsn,Cirigliano:2005ck}.
The flavor group is 
\begin{align}
G_F =& SU(3)_Q\otimes SU(3)_u \otimes SU(3)_d \nonumber \\
	&\otimes SU(3)_L \otimes SU(3)_e 
\otimes SU(3)_\nu, 
\end{align}
where we have allowed for the existence of three generations of sterile neutrinos. The fermions
are in the fundamental or anti-fundamental representation, as
\begin{align}
	Q \sim (\bm 3, \bm 1, \bm 1, \bm 1, \bm 1, \bm 1), \qquad 
	u^c \sim (\bm 1 , \bm{\overline{ 3}}, \bm 1, \bm 1, \bm 1, \bm 1),  \nonumber \\
	d^c \sim (\bm 1, \bm 1, \bm{\overline{3}}, \bm 1, \bm 1, \bm 1), \,\,\qquad 
	L \sim (\bm 1, \bm 1, \bm 1, \bm 3, \bm 1, \bm 1),  \nonumber \\
	e^c \sim (\bm 1, \bm 1, \bm 1, \bm 1, \bm{\overline{3}}, \bm 1), \qquad 
	\nu^c \sim (\bm 1, \bm 1, \bm 1, \bm 1, \bm 1, \bm{\overline{3}}). 
\end{align}
The Yukawa terms and the Majorana mass term explicitly break the flavor symmetry. In order 
to preserve the symmetry, the Yukawa couplings and the Majorana mass term are promoted to
spurion fields. The requirement of MFV is that all flavor breaking interactions should appear 
in the same pattern as for the dimension-4 SM. 
The Yukawa terms and the Majorana mass term take the form
\begin{align}
	-\Lagr_{\textrm{spurion}} =& Y_u Q H u^c + Y_d Q H^* d^c + Y_e L H^* e^c \nonumber \\
	& + Y_\nu L H \nu^c + \frac{1}{2}M_\nu \nu^c \nu^c + \textrm{h.c.}
	\label{eq:neutrinoMass}
\end{align}
The spurion fields transform as 
\begin{align}
	&Y_u \sim (\bm{\overline{3}}, \bm 3, \bm 1, \bm 1, \bm 1, \bm 1), \qquad
	Y_d \sim (\bm{\overline{3}}, \bm 1, \bm 3, \bm 1, \bm 1, \bm 1), \qquad \nonumber \\
	&Y_e \sim (\bm 1, \bm 1, \bm 1, \bm{\overline{3}}, \bm 3, \bm 1), \qquad  
	Y_\nu \sim (\bm 1, \bm 1, \bm 1, \bm{\overline{3}}, \bm 1, \bm{3}), \qquad \nonumber \\	
	&M_\nu \sim (\bm 1, \bm 1, \bm 1, \bm 1, \bm 1, \bm 3\otimes \bm 3). \qquad 	
\end{align}
For an operator to be invariant under the MFV group, the following relations must hold,
\begin{align}
	\label{eq:MFVeq1}
	\frac{1}{3}\left(N_Q + N_{Y_u^\dagger} + N_{Y_d^\dagger}\right)\qquad\qquad& \nonumber\\
	- \frac{1}{3}\left(N_{Q^\dagger} + N_{Y_u} + N_{Y_d} \right) &\in \Z, \\ 
	\label{eq:MFVeq2}
	\frac{1}{3}\left(N_{u^\dagger} + N_{Y_u} \right) 
	- \frac{1}{3}\left(N_{u} + N_{Y_u^\dagger}  \right) &\in \Z, \\
	\label{eq:MFVeq3}
	\frac{1}{3}\left(N_{d^\dagger} + N_{Y_d} \right) 
	- \frac{1}{3}\left(N_{d} + N_{Y_d^\dagger}  \right) &\in \Z, \\
	\label{eq:MFVeq4}
	\frac{1}{3}\left(N_{L} + N_{Y_e^\dagger} + N_{Y_\nu^\dagger} \right)\qquad\qquad& \nonumber \\
	- \frac{1}{3}\left(N_{L^\dagger} + N_{Y_e} + N_{Y_\nu} \right) &\in \Z, \\
	\label{eq:MFVeq5}
	\frac{1}{3}\left(N_{e^\dagger} + N_{Y_e} \right) 
	- \frac{1}{3}\left(N_{e} + N_{Y_e^\dagger}  \right) &\in \Z, \\
	\label{eq:MFVeq6}
	\frac{1}{3}\left(N_{\nu^\dagger} + N_{Y_\nu} \right) 
	- \frac{1}{3}\left(N_{\nu} + N_{Y_{\nu}^\dagger}  \right) &\nonumber \\
	+ \frac{2}{3}\left(N_{M_\nu} - N_{M_\nu^\dagger} \right)
	&\in \Z. 
\end{align}
By summing Eqs.~\eqref{eq:MFVeq1}-\eqref{eq:MFVeq3}, we find that baryon number must be an 
integer, which already followed from hypercharge invariance (or invariance under $SU(3)_c$).
Adding Eqs.~\eqref{eq:MFVeq4}-\eqref{eq:MFVeq6}, we have that
\begin{align}
	\frac{1}{3}\left(N_L + N_{e^\dagger} + N_{\nu^\dagger} \right) 
	&- \frac{1}{3}\left( N_{L^\dagger} + N_{e} + N_{\nu} \right) \nonumber \\
	&+ \frac{2}{3}\left( N_{M_\nu} - N_{M_\nu^\dagger} \right)  \in \Z.
\end{align}
Using the definition of lepton number, Eq.~\eqref{eq:DeltaL}, we have that
\begin{align}
	\label{eq:3DeltaLwithMnu}
	\frac{1}{3}\Delta L + \frac{2}{3}\left( N_{M_\nu} - N_{M_\nu^\dagger}\right) \in \Z.
\end{align}
In the case where $N_{M_\nu} = N_{M_\nu^\dagger}$, we find agreement with Eq.~\eqref{eq:DeltaL3}.
The difference between Eqs.~\eqref{eq:DeltaL3} and \eqref{eq:3DeltaLwithMnu} is due to the 
inclusion of the sterile neutrinos, which explicitly break the flavor symmetry via the Majorana
mass term and Yukawa interaction. Also, the Majorana mass term violates lepton number
by two units.

\section*{Proton Decay}
The group-theoretical considerations presented above have phenomenological consequences.
Experimentally relevant are the implications for the search for proton decay.
Cherenkov-radiation detectors like Super-Kamiokande are used to search for certain
potential decay channels of the proton \cite{Fukuda:2002uc}.

Baryon number and lepton number are accidental symmetries of the SM, and are violated
in many grand unified theories \cite{Nath:2006ut,Pati:1974yy,Lee:1994vp,Shaban:1992he,Ellis:2002vk,Kim:2002im,Pati:2003qia,Buchmuller:2004eg} (see Ref.~\cite{Grinstein:2006cg} for a discussion on MFV in grand unified theories). In many of the 
beyond SM theories, the dominant decay channel of the proton is 
$p\rightarrow e^+ \pi^0$ (or $p\rightarrow \mu^+ \pi^0$), where the proton $p$ decays
into a charged anti-lepton and a neutral pion. The neutral pion would decay further
to two photons, which could be detected by the Cherenkov-radiation detector. 
From an effective-field-theory perspective, the two-body decay of the proton could
arise from a dimension-6 operator $qqql$ \cite{Weinberg:1979sa,Wilczek:1979hc}.
The null results from the Super-Kamiokande experiment have pushed the scale of 
new physics associated with the dimension-6 operator $qqql$ to $\Lambda \sim 10^{15}$ GeV.
This corresponds to a bound on the partial life-time of the proton of 
$\tau_{N\rightarrow M l} \geq 10^{34}$ years \cite{Nishino:2012bnw,Miura:2016krn}.

However, with the presence of certain flavor symmetries and with no sterile neutrinos, 
the dimension-6 operators
resulting in proton decay are excluded. 
The baryon-number violating 
operators with lowest mass dimension have mass dimension 9. Thus, we need to analyze the decay channels resulting
from the new leading baryon-number violating operators.

The only dimension-9 operators with $\Delta B=\Delta L/3=1$ are
$u^{c\dagger}u^{c\dagger}u^{c\dagger}e^{c\dagger}LL$ and 
$u^{c\dagger}u^{c\dagger}QLLL$. However, neither contributes to 
three-body nucleon decay at tree-level since both contain heavier quarks, e.g., a charm or top quark \cite{Weinberg:1980bf}.
At dimension-10, the operator $d^{c\dagger}d^{c\dagger}d^{c\dagger}L^{\dagger} L^{\dagger} L^{\dagger} H^\dagger$, with 
$\Delta B= -\Delta L/3=1$, could contribute to nucleon decay, through a four-body decay \cite{Weinberg:1980bf}.
The lowest-dimensional operators with $\Delta B=\Delta L/3=1$ which contributes to three-body proton decay
at tree-level are dimension-11 operators, such as $u^{c\dagger}d^{c\dagger}QLLLHH$ \cite{Weinberg:1980bf}.
%
If one posits the existence of flavor symmetries at high scales, then it may be very likely
that the dominant contribution to proton decay would come from such higher-dimensional
operators.
This could result in a three-body decay, with three leptons in the final state. 
The estimated decay width is 
\begin{align}
	\Gamma \sim \frac{1}{512\pi^3}\left(\frac{\langle H\rangle^2}{\Lambda^7}\right)^{2} \Lambda_{\textrm{QCD}}^{11},
\end{align}
where $\Lambda$ is the scale associated with the intermediate flavor interaction and
$\langle H\rangle$ is the vacuum expectation value of the Higgs field.
Current experiments would be sensitive to effects from these operators if 
the scale is $\Lambda \sim 10^5$ GeV.

Some searches for three-lepton decays of the proton have been performed, 
but not exhaustively across all possible decay channels \cite{Takhistov:2014pfw}. 
Since, on very general grounds, these three-lepton decay channels may be a positive 
indication of an intermediate scale associated with flavor, 
further experimental investigation would be valuable.

\section*{Majorana masses}
Next we consider Majorana mass terms. By excluding the sterile neutrinos in the 
dimension-4 SM, we ask whether higher-dimensional operators
resulting in Majorana mass terms for the SM neutrinos are allowed.
From the discussion on MFV, by setting $N_{M_\nu}=N_{M_\nu^\dagger}=0$, we find that 
lepton number can only be violated in multiples of 3, Eq.~\eqref{eq:DeltaL3}.
This is an explicit example of a general result that, excluding fermions in the adjoint 
representation of $SU(2)$, no neutrino mass term is allowed.
That is, if one wants to generate Majorana neutrino mass terms,
and have a certain flavor symmetry, only two options are available. One could either have the
leptons be in the adjoint representation of a flavor $SU(2)$ group (see e.g. Ref.~\cite{Berryman:2016mfg}), or introduce
some explicit violation of the flavor symmetry, e.g., as in Eq.~\eqref{eq:neutrinoMass}.


\begin{acknowledgements}
We are grateful to A. de Gouv\^{e}a and B. Grinstein for comments on the manuscript.
The work of AH was supported in part by the Danish National Research Foundation (DNRF91) and the Carlsberg Foundation.
We thank the Department of Physics,  UC San Diego for the generous hospitality when part of this work was completed.
\end{acknowledgements}

\bibliography{baryonnumber}{}

\end{document}